\documentstyle[sprocl]{article}

\def\be{\begin{equation}}
\def\ee{\end{equation}}
\def\bea{\begin{eqnarray}}
\def\eea{\end{eqnarray}}

\begin{document}

\title{Parity Violating Measurements of Neutron Densities: Implications for Neutron Stars}

\author{C. J. Horowitz}
\address{Dept. of Physics and Nuclear Theory Center,
Indiana University, Bloomington, IN 47405 USA \\ E-mail: horowitz@iucf.indiana.edu}

\author{J.  Piekarewicz}

\address{Department of Physics, Florida State University, 
Tallahassee, FL 32306 \\ E-mail: jorgep@scri.fsu.edu}

\maketitle\abstracts{ 
Parity violating electron scattering can measure the neutron density of a heavy nucleus accurately and model independently.  This is because the weak charge of the neutron is much larger then that of the proton.   The Parity Radius Experiment (PREX) at Jefferson Laboratory aims to measure the root mean square neutron radius of $^{208}$Pb with an absolute accuracy of 1\% ($\pm 0.05$ Fm).  This is more accurate then past measurements with hadronic probes, which all suffer from controversial strong interaction uncertainties.  PREX should clearly resolve the neutron-rich skin.  Furthermore, this benchmark value for $^{208}$Pb will provide a calibration for hadronic probes, such as proton scattering, which can then be used to measure neutron densities of many exotic nuclei.  The PREX result will also have many implications for neutron stars.  The neutron radius of Pb depends on the pressure of neutron-rich matter: the greater the pressure, the larger the radius as neutrons are pushed out against surface tension. The same pressure supports a neutron star against gravity.  The Pb radius is sensitive to the equation of state at normal densities while the radius of a 1.4 solar mass neutron star also depends on the equation of state at higher densities.  Measurements of the radii of a number of isolated neutron stars such as Geminga and RX J185635-3754 should soon improve significantly. By comparing the equation of state information from the radii of both Pb and neutron stars one can search for a softening of the high density equation of state from a phase transition to an exotic state.  Possibilities include kaon condensates, strange quark matter or color superconductors.  
}

\section{Introduction}

The size of a heavy nucleus is one of its most basic properties.  However,
because of a neutron skin of uncertain thickness, the size does not follow from
measured charge radii and is relatively poorly known.  For example, the root
mean square neutron radius in $^{208}$Pb, $R_n$ is thought to be about 0.2 Fm
larger then the proton radius $R_p\approx 5.45$ Fm.  An accurate measurement of
$R_n$ would provide the first clean observation of the neutron skin in a stable
heavy nucleus.  This is thought to be an important feature of all heavy nuclei.

Ground state charge densities have been determined from elastic electron
scattering, see for example ref.\cite{Sick}.  Because the densities are both
accurate and model independent they have had a great and lasting impact on
nuclear physics.  They are, quite literally, our modern picture of the nucleus.

In this paper we discuss future parity violating measurements of neutron densities.
These purely electro-weak experiments follow in the same tradition and can be
both {\it accurate} and {\it model independent}.  
Neutron density measurements have implications for nuclear structure, atomic
parity nonconservation (PNC) experiments, isovector interactions, the structure
of neutron rich radioactive beams, and neutron rich matter in astrophysics.  It
is remarkable that a single measurement has so many applications in atomic,
nuclear and astrophysics.

Donnelly, Dubach and Sick\cite{donnelly} suggested that parity violating
electron scattering can measure neutron densities.  This is because the
$Z-$boson couples primarily to the neutron at low $Q^2$.  Therefore one can
deduce the weak-charge density and the closely related neutron density from
measurements of the parity-violating asymmetry in polarized elastic scattering.

Of course the parity violating asymmetry is very small, of order a part per
million.  Therefore measurements were very difficult.  However, a great deal of
experimental progress has been made since the Donnelly {\it et. al.}  
suggestion, and since the early SLAC experiment \cite{SLAC}. 
This includes the Bates $^{12}$C experiment\cite{carbon12}, 
Mainz $^{9}$Be experiment \cite{Heil}, SAMPLE \cite{sample1} 
and HAPPEX \cite{happex1}.  The relative speed of the HAPPEX result and the very 
good helicity correlated beam properties of CEBAF show that very accurate 
parity violation measurements are possible.  Parity violation is now an 
established and powerful tool.

It is important to test the Standard Model at low energies with atomic parity nonconservation (PNC), see for example the Colorado measurement in Cs \cite{Wieman98,Wieman99}.  These
experiments can be sensitive to new parity violating interactions such as
additional heavy $Z-$bosons.  Furthermore, by comparing atomic PNC to higher
$Q^2$ measurements, for example at the $Z$\ pole, one can study the momentum 
dependence of Standard model radiative corrections.  However, as the 
accuracy of atomic PNC experiments improves they
will require increasingly precise information on neutron
densities\cite{Pollock,Chen}.  
This is because the parity violating interaction is
proportional to the overlap between electrons and neutrons.  In the future the
most precise low energy Standard Model test may involve the combination of an
atomic PNC measurement and parity violating electron scattering to constrain the
neutron density.

There have been many measurements of neutron densities with strongly 
interacting probes such as pion or proton elastic scattering, see for 
example ref. \cite{Ray}.  Unfortunately, all such
measurements suffer from potentially serious theoretical systematic errors.  As
a result no hadronic measurement of neutron densities has been generally
accepted by the field.  

Relative measurements of isotope differences of neutron radii can be more accurate.  See for example \cite{Ray2}.  Therefore one can use a single parity violating measurement of the neutron radius of $^{208}$Pb to ``calibrate" hadronic probes.  Then these hadronic probes can be used to measure neutron radii of many other stable and unstable nuclei.  For example, ($^3$He,T) measurements of neutron radii differences for Sn isotopes were based on measuring spin dipole strength and a spin dipole sum rule along with assuming a theoretical Hartree Fock radius for $^{120}$Sn \cite{RCNP}.

Finally, there is an important complementarity between neutron radius
measurements in a finite nucleus and measurements of the neutron radius of a
neutron star.  Both provide information on the equation of state (EOS) of dense
matter.  In a nucleus, $R_n$ is sensitive to the EOS at normal nuclear densities.
The neutron star radius depends on the EOS at higher densities.  In the near future, we expect a number of improving radius measurements for nearby
isolated neutron stars such as Geminga \cite{Geminga} and RX J185635-3754 \cite{Nature}.

We now present general considerations for neutron density measurements, discusses 
possible theoretical corrections, outline an approved Jefferson Laboratory experiment on $^{208}$Pb and then relate this Pb measurement to ongoing measurements of neutron star radii.

\section{General Considerations}

In this section we illustrate how parity violating electron scattering
measures the neutron density and discuss the effects of Coulomb
distortions and other corrections.  These corrections are either small or well known so the interpretation of a measurement is clean.

\subsection{Born Approximation Asymmetry}

The weak interaction can be isolated by measuring the parity-violating asymmetry in the cross section for the scattering of left (right) handed electrons.  In Born approximation the parity-violating asymmetry is,
\begin{equation}A_{LR}=\frac{G_FQ^2}{4\pi\alpha\sqrt{2}}
\Biggl[4\sin^2\theta_W-1+\frac{F_n(Q^2)}{F_p(Q^2)}\Biggr],
\label{equation_bornasy}
\end{equation} 
with $G_F$ the Fermi constant and $\theta_W$ the weak mixing angle.  The Fourier transform of the proton distribution is $F_p(Q^2)$, while that of the neutron distribution is $F_n(Q^2)$, and $Q$ is the momentum transfer.  The
asymmetry is proportional to $G_F Q^2/\alpha$ which is
just the ratio of $Z^0$ to photon propagators. 
Since 1-4sin$^2\theta_W$ is small and $F_p(Q^2)$ is known we see that $A_{LR}$ directly measures $F_n(Q^2)$.  Therefore, $A_{LR}$ provides a practical method to cleanly measure the
neutron form factor and hence $R_n$.

\subsection{Coulomb distortions}

By far the largest known correction to the asymmetry comes from coulomb
distortions.  By coulomb distortions we mean repeated electromagnetic
interactions with the nucleus remaining in its ground state.  All of the $Z$
protons in a nucleus can contribute coherently so distortion corrections are
expected to be of order $Z\alpha/\pi$.  This is 20 \% for ${}^{208}$Pb.

Distortion corrections have been accurately calculated in ref. \cite{cjh}.  Here
the Dirac equation was numerically solved for an electron moving in a coulomb
and axial-vector weak potentials.  From the phase shifts, all of the elastic
scattering observables including the asymmetry can be calculated.

Other theoretical corrections from meson exchange currents, parity admixtures in the ground state, dispersion corrections, the neutron electric form factor, strange quarks, the dependence of the extracted radius on the surface shape, etc. are discussed in reference \cite{bigpaper}.  These are all small.  Therefore the interpretation of a parity violating measurement is very clean.

\section{Parity Radius experiment}

The Parity Radius Experiment (P-ReX) will measure the parity violating asymmetry for elastic electron scattering from $^{208}$Pb \cite{PREX}.  This Jefferson Laboratory Hall A experiment will use 850 MeV electrons scattered at six degrees.  The planned 3\% accuracy in the approximately 0.7 parts per million asymmetry will allow one to deduce the neutron root mean square radius $R_n$ to 1\% ($\approx \pm 0.05$ Fm).  The neutron radius $R_n$ is expected to be about 0 to 0.3 Fm larger then the proton radius $R_p$.  Therefore PREX should cleanly resolve the neutron skin $R_n-R_p$.

The target will be a thick foil, enriched in $^{208}$Pb, sandwiched between two thin diamond foils.  The very high thermal conductivity of the diamond keeps the Pb from melting and allows a high beam current of order 100 microamps.  Note, the thin diamond foils introduce only a few percent background.  Furthermore, the asymmetry from $^{12}$C can be calculated with high accuracy so this background is not a problem for the interpretation of the experiment.  

PREX requires some improvements in the helicity correlated beam properties and an improvement in the measurement of the absolute beam polarization in Hall A.  This is presently of order 3\% and needs to be improved to 1-2\%.  It should take about 30 days of beam time to get the 3\% statistics.    

\section{Implications of the $^{208}$Pb neutron radius for neutron stars}

It is an exciting time to study neutron stars.  These gigantic atomic nuclei are more massive then the Sun and yet have a radius of only about 10 kilometers.  New telescopes, operating at many different wave lengths, are finally turning these theoretical curiosities into detailed observable worlds.  The structure of a neutron star depends only on the equation of state (EOS) of neutron rich matter together with the know equations of General Relativity.  The equation of state gives the pressure as a function of (energy) density.  Densities in neutron stars are comparable to, or greater, then the densities in atomic nuclei.  The central density of a 1.4 solar mass neutron star is expected to be a few times greater then the saturation density of nuclear matter, $\rho_0\approx 0.16$ nucleons per Fm$^3$.

Likewise the neutron radius of a conventional atomic nucleus such as $^{208}$Pb also depends on the equation of state of neutron rich matter.  Higher pressures lead to greater neutron radii and thicker neutron skins as neutrons are pushed out against surface tension.  Indeed, Alex Brown finds a strong correlation between the pressure of neutron matter at $\rho\approx 0.1$ Fm$^{-3}$ and the neutron radius in $^{208}$Pb \cite{Brown}.  This correlation is valid for many different nonrelativistic and relativistic effective interactions.  The density $\rho=0.1$ Fm$^{-3}$ is about 2/3 of $\rho_0$ and represents some average over the interior and surface density of the nucleus.

Therefore, the neutron radius in Pb has many implications for the structure of neutron stars and several other areas of astrophysics.  The common unknown is the equation of state of neutron rich matter.  Information on the EOS from a measurement of $R_n$ for $^{208}$Pb could be very important for astrophysics.

\subsection{Neutron Skin versus Neutron Star Crust}

The Pb radius constrains the EOS at normal densities $\approx 0.1$ Fm$^{-3}$.  Neutron stars are expected to undergo a phase transition near this density from a solid crust to a liquid interior.  We have shown that the Pb radius is strongly correlated with the liquid to solid transition density \cite{Skins}.  A high pressure for neutron rich matter more quickly favors the uniform liquid over the nonuniform solid.  Therefore, a large neutron radius in Pb implies a low transition density for the crust.  Thus, a measurement of the thickness of the neutron rich skin in Pb helps determine the thickness of the solid crust of a neutron star.  We note that both the skin of a heavy nucleus and the crust of a neutron star are made of neutron rich matter at similar densities.  Many neutron star observables such as glitches in the rotational period, gravitational waves from quadrupole deformations, and the surface temperature depend on the thickness of the crust.  Adrian Cho has written a short popular article on using this pression measurement in Pb to learn about neutron star crusts \cite{Cho}.    

\subsection{Pb radius versus Neutron Star Radius}

In general the radius of a neutron star $R_*$ is related to the neutron radius in Pb $R_n$.  A large $R_n$ implies a high pressure for the EOS of neutron rich matter and this same pressure supports a star against gravity.  Therefore, a larger $R_n$ might imply a larger $R_*$.  However, the radius of a neutron star $R_*$ depends on the EOS of neutron rich matter over a range of densities from near $\rho_0$ to higher densities.  In contrast, $R_n$ only depends on the EOS at $\rho_0$ and lower densities.  Thus, $R_n$ only constrains the low density EOS.  Models with different high density behavior can have the same $R_n$ but predict different $R_*$.  Therefore, we find no unique relationship between $R_*$ and $R_n$ \cite{Radii}.

One way to characterize the different information on the EOS contained in $R_n$ compared to $R_*$ is to consider low mass neutron stars.  Most, well measured, neutron stars have masses near 1.4 solar masses.  These stars have central densities significantly above $\rho_0$.  Instead, 0.5 Solar mass neutron stars are expected to have central densities only slightly greater then $\rho_0$.  We find a sharp correlation between $R_*$ for 0.5 solar mass neutron stars and $R_n$ \cite{smallNS}.  This is because, now, both $R_*$ and $R_n$ depend on the EOS at similar densities.         

Note, such low mass neutron stars probably don't exist.  This is because conventional stars with cores near 0.5 solar masses are not expected to collapse.  Thus, $R_n$ contains unique information on the low density EOS that can not be obtained by measuring neutron star radii directly.  Furthermore, {\it measuring both $R_n$ in Pb and $R_*$ for a neutron star \footnote{That is probably near 1.4 solar masses.} provides important information on the density dependence of the EOS}.  

For example, if $R_n$ is measured to be relatively large, this implies a stiff (high pressure) EOS at normal nuclear densities.  If $R_*$ is also measured to be relatively small, say near 10 km, this implies a soft (low pressure) high density EOS.  This softening of the EOS with increasing density could be strong evidence for a phase transition of neutron rich matter to some exotic phase.  Note, an exotic phase that increases the pressure would not be thermodynamicly favored.  
There is much speculation on possible high density exotic phases for neutron rich matter.  Examples include: kaon condensation, strange quark matter, or color superconductivity.  

Neutron stars provide essentially the only way to study cold very dense matter.  Relativistic heavy ion collisions can reach high energy densities but not at low temperatures.  Thus, if an exotic phase exits at high density it is very hard to find experimental evidence.  This comparison of the EOS information from $R_n$ and $R_*$ may be one of the few sharp signals.  Therefore, it is important to measure {\it both} the neutron radius of a heavy nucleus $R_n$, and the radius of a neutron star $R_*$.  

\subsection{Measurements of Neutron Star Radii}

There are several ongoing measurements of the radius of neutron stars.  Most of these are based on measuring the stars luminosity, distance and surface temperature $T$.  If the star were a black body, the luminosity would be $\sigma T^4$ times the surface area.  Thus, the surface area $4\pi R_\infty^2$ and effective radius $R_\infty$ can be deduced.  Corrections for non black body behavior can be made with model atmospheres.  

Because of the curvature of space in the Star's very strong gravitational field, the effective radius $R_\infty$ is somewhat larger then the coordinate radius $R_*$,
\begin{equation}
R_\infty=R_*/(1-2GM/R_*)^{1/2},
\end{equation}
where $M$ is the Star's mass and $G$ is Newton's constant.  Some of the light from the far side of the star is bent by gravity and still reaches an observer.  This makes the star appear larger.  Note, the light is also gravitationally red shifted so that the actual surface temperature is about 30\% higher then the apparent temperature deduced from the observed spectrum.      

The Stony Brook group has fit the visible and X-ray spectrum of the isolated nearby neutron star RX J185635-3754 with a model Fe atmosphere \cite{Walter}.  Their fit combined with a preliminary parallax distance to RX J185635 of 61 parsecs (about 180 light years) yields a very small radius of $R_*=6$ km!  This radius is smaller then that predicted by any present neutron matter EOS and seems unrealistic.  However Kaplan et al. \cite{Kaplan} have questioned the parallax distance.  They reanalyze the same Hubble optical images and infer a larger distance.  For their distance, $R_\infty=15\pm 6$ km.  Note for a 1.4 solar mass star, $R_\infty=15$ km corresponds to $R_*\approx 13$ km.  This new value is fully consistent with many neutron matter EOSs.  However, the error is still large.

Sanwal et al. \cite{Vela} fit the X-ray spectrum of the well known Vela pulsar with a high energy power law, from the pulsar mechanism, and a thermal component from a magnetized hydrogen atmosphere.  They deduce $R_\infty=15.5\pm 1.5$ km assuming a distance to Vela of 250 parsec.  Finally Rutledge et al. analyze the X-ray spectra of a neutron star in the globular star cluster NGC 5139, for which the distance to the cluster is well known \cite{Rutledge}.  Their preliminary result is $R_\infty=14.3\pm 2.5$ km.

In the near future, we should have better distance measurements to RX J185635 and better X-ray spectra.  Also, there will be better measurements on other nearby isolated neutron stars such as Geminga, and more measurements of neutron stars in globular clusters.  This will allow checks on neutron star radii measured for stars with different surface temperatures and magnetic field strengths.  A reasonable near term goal is a number of neutron star radius measurements accurate to about one km.

\section{Conclusion}

With the advent of high quality electron beam facilities
such as CEBAF, experiments for accurately measuring the
weak density in nuclei through parity violating
elastic electron scattering (PVES) are feasible.
From parity violating asymmetry measurements, 
one can extract the neutron density of a heavy nucleus accurately and model independently.  This is because the weak charge of a neutron is much larger then that of a proton.  Therefore, the $Z^0$ boson couples primarily to neutrons (at low momentum transfers).

These neutron density measurements allow a direct test of mean field theories and other models of the size and shape of nuclei.  They can have a fundamental and lasting impact on nuclear physics.  Furthermore, PVES measurements have important implications for atomic parity nonconservation (PNC) experiments.  Atomic PNC measures the overlap of atomic electrons with neutrons.  High precsion PNC experiments will need accurate neutron densities.  In the future, it may be possible to combine atomic PNC experiments and PVES to provide a precise test of the Standard Model at low energies.

The Parity Radius Experiment at Jefferson Laboratory aims to measure the neutron radius in $^{208}$Pb to 1\% with parity violating elastic electrons scattering.  This will provide unique information on the equation of state (EOS) of neutron rich matter at normal nuclear densities.  The EOS describes how the pressure depends on the density.  This information has many astrophysical implications.

The structure of a neutron star depends only on the EOS.  There are many ongoing measurements of the radius of neutron stars.  These are sensitive to the EOS at greater then nuclear densities.  By comparing the EOS information from the $^{208}$Pb neutron radius measurement with that from neutron star measurements one can deduce the density dependence of the EOS.  This allows one to search for a softening (lower pressure) of the high density EOS from a possible phase transition to an exotic phase for neutron rich matter.  Possible phases include kaon condensates, strange quark matter and color superconductors.

\section*{Acknowledgments}
The work on parity violating measurements of neutron densities was done in collaboration with Robert Michaels, Steven Pollock and Paul Souder.  We acknowledge financial support from DOE grants: DE-FG02-87ER40365 and DE-FG05-92ER40750.

\end{document}